\documentclass[10pt, conference, compsocconf]{IEEEtran}
\pdfoutput=1
\usepackage{ifpdf}
\usepackage{cite}
\usepackage{url}

\ifCLASSINFOpdf
  \usepackage[pdftex]{graphicx}
\else
\fi

\usepackage[cmex10]{amsmath}
\usepackage{algorithmic}
\usepackage{array}
\usepackage{url}
\usepackage[textwidth=1cm]{todonotes}

\begin{document}
%
\title{Technical report:\\A Decentralized Parallelization-in-Time Approach with Parareal}


\author{

\IEEEauthorblockN{Martin Schreiber}
\IEEEauthorblockA{
University of Exeter\\
CEMPS\\
United Kingdom\\
M.Schreiber@exeter.ac.uk}

\and
\IEEEauthorblockN{Adam Peddle}
\IEEEauthorblockA{
University of Exeter\\
CEMPS\\
United Kingdom\\
ap553@exeter.ac.uk}

\and
\IEEEauthorblockN{Terry Haut}
\IEEEauthorblockA{Los Alamos National Laboratory\\
CNLS, MS B258\\
Los Alamos\\
terryhaut@gmail.com}

\and
\IEEEauthorblockN{Beth Wingate}
\IEEEauthorblockA{
University of Exeter\\
CEMPS\\
United Kingdom\\
B.Wingate@exeter.ac.uk}

}


%


\maketitle

\begin{abstract}
With steadily increasing parallelism for high-performance architectures, simulations requiring a good strong scalability are prone to be limited in scalability with standard spatial-decomposition strategies at a certain amount of parallel processors.
This can be a show-stopper if the simulation results have to be computed with wallclock time restrictions (e.g.\,for weather forecasts) or as fast as possible (e.g. for urgent computing).
Here, the time-dimension is the only one left for parallelization and we focus on Parareal as one particular parallelization-in-time method.

We discuss a software approach for making Parareal parallelization transparent for application developers, hence allowing fast prototyping for Parareal.
Further, we introduce a decentralized Parareal which results in autonomous simulation instances which only require communicating with the previous and next simulation instances, hence with strong locality for communication.
This concept is evaluated by a prototypical solver for the rotational shallow-water equations which we use as a representative black-box solver.

\end{abstract}

\begin{IEEEkeywords}
  high-performance computing, parallelization in time, parareal, decentralized
\end{IEEEkeywords}

%
\IEEEpeerreviewmaketitle

\section{Introduction}

Over the last decade an improvement in the performance of simulations executed on 
supercomputers has been accomplished by increasing the number of parallel
data processing pipelines on the core as well as on the instruction level.
This is in contrast to previous decades where performance was mainly improved
through increasing the CPU's clock rate which nowadays almost stagnated (see \cite{kogge2011using}).
This recent type of architectural development has a significant impact on strong scaling problems:
the
spatial decomposition is at one point dominated by the communication latencies at a
fixed number of processors and no further improvement in performance
 can be achieved through the utilization of more computing cores. In
combination with MPI-related restrictions, using more cores can even
lead to less performance. Since the trend of increasing supercomputer
performance through more data parallelisation is likely to continue, this
will have a significant impact on the future of HPC applications and
in particular for problems with strong scaling.
In this paper, we focus on
simulations with run-time requirements such as sub-realtime (for
weather and climate, e.g.).
With the aforementioned tendency to increase
the number of parallel data processing pipelines, this requires
exploiting new ways of parallelisation, including those gained by 
using insights from novel mathematical formulations.

\begin{figure}
	\center
	\includegraphics[width=0.45\textwidth]{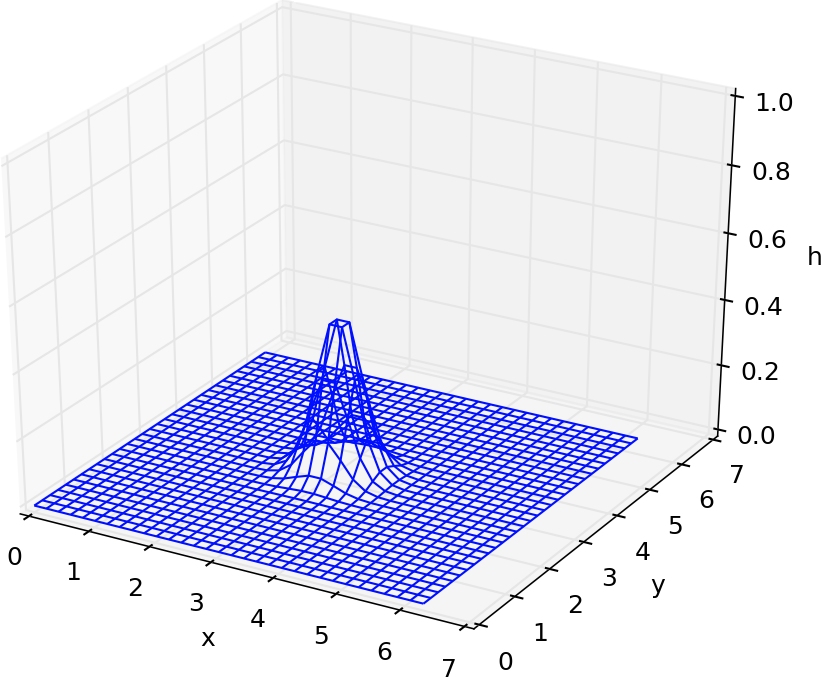}\\
	\includegraphics[width=0.45\textwidth]{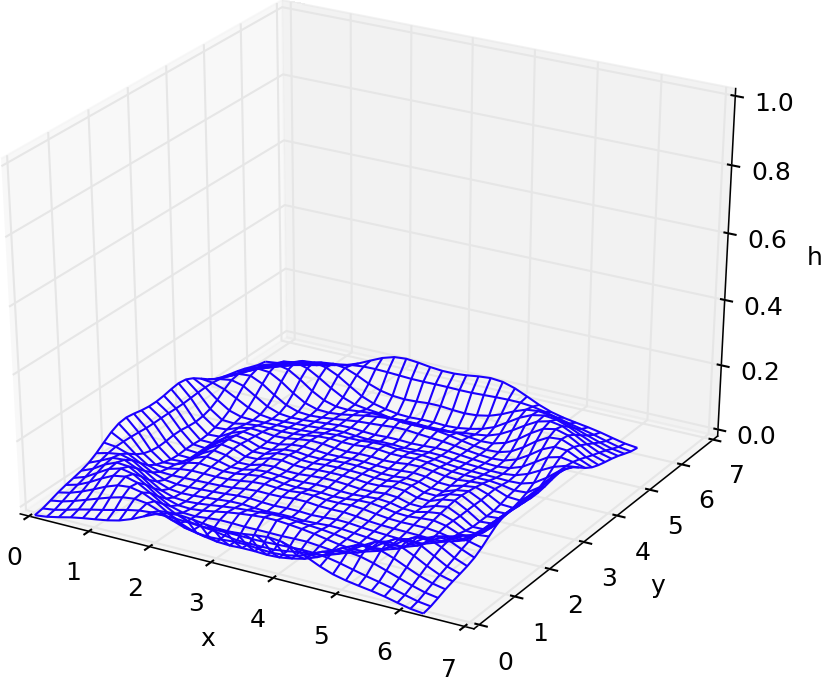}
	\caption{Water surface height of the initial condition given by a Gaussian distribution (top image) and selected solutions of the rotational shallow water equations (RSWE) which we solve in this work (bottom image).}
\end{figure}

Our work is based on the parallel-in-time iterative method called 'Parareal'
\cite{lions2001resolution}.  Here, the simulation time is divided into
\emph{coarse time intervals}.  Two different propagators are used: a
fine propagator (also the default for standard space-parallelisation
methods) and a coarse propagator, which has to be of lower complexity
than the fine propagator over the coarse time interval.  A
\emph{coarse propagator} (approximation) is used to compute an
approximation of the solution at the start of each coarse time
interval. 
The Parareal parallel-in-time method then uses the coarse
propagator to estimate solutions at the end of the coarse time
intervals.  This is followed by a combination of fine and coarse
propagators in each coarse time interval and is used as an iterative
method to improve the approximated solution.  This can be executed
massively parallel since computations on all intervals are independent in space.
Such an approach can be implemented event-based with a dynamic task
scheduling library\cite{elwasif2011dependency} or with a
centralized manager-worker task distribution\cite{Aubanel2011172}.
However, the potentials of the locality properties of the data flow
with the Parareal method were so far not considered in these works.
In this work we use a black-box solver which represents an
experimental implementation of the rotating shallow water
equations (RSWE).

\section{Parareal}
\label{sec:parareal_algorithm}

\begin{figure}
\center
\includegraphics[width=0.47\textwidth]{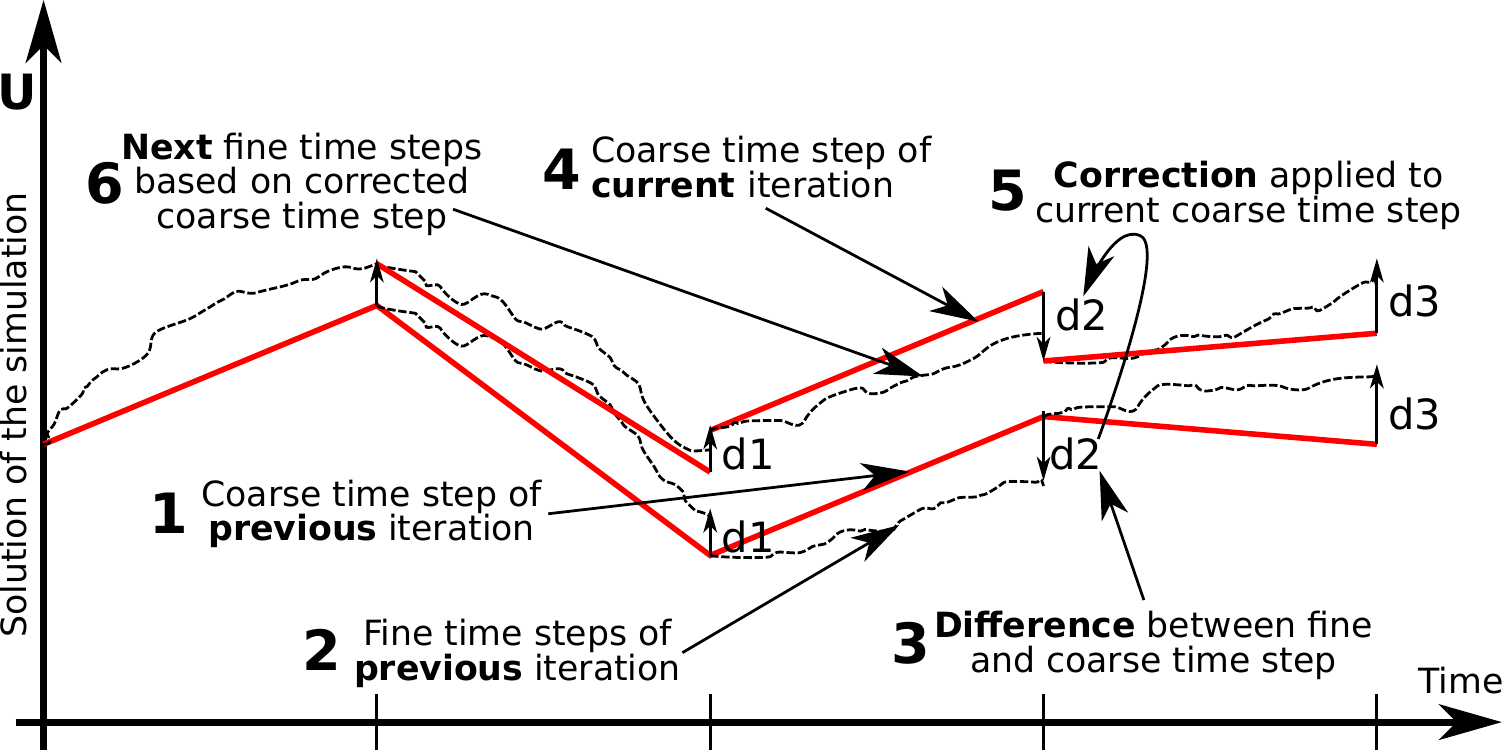}
\caption{Sketch of the Parareal algorithm for \emph{second iteration} with the focus on the third coarse time interval:
After computing the coarse timestep (1) and the fine time stepping (2) as part of the first iteration, the difference is computed and buffered (3).
Then, the next coarse time step is executed based on updated initial values for this coarse time interval (4).
The result of this coarse time step is then corrected with the previously computed difference (5).
\label{fig:sketch_parareal}}
\end{figure}

Here, we give an overview of the Parareal algorithm which is employed in this work for the parallelization-in-time.
This algorithm was initially presented in \cite{lions2001resolution}, but has its roots in earlier works by \cite{Nievergelt_64}. Particular attention has been paid to the parallel implementation of the Parareal method by \cite{Maday_Turinici_05}. For a more detailed review of the history of the Parareal method, the reader is referred to \cite{Gander_15}.
A sketch of the algorithm for ODEs is given in Fig.\,\ref{fig:sketch_parareal}.

Consider some general system of Partial Differential Equations of the form:
\begin{equation}
  \frac{\mathrm{d}\mathbf{U}}{\mathrm{d}t} = \mathbf{f}(\mathbf{U})\text{,   }\mathbf{U}(0) = \mathbf{U}_{0}\text{,   }t\in [0,T].
\end{equation}
Where $\mathbf{f}(\mathbf{U})$ is some differential operator which is not necessarily linear.
The Parareal algorithm is defined by two propagation operators over the $[t_{n}, t_{n+1}]$ time interval, $\mathcal{G}(t_{n},\mathbf{U}_{n})$, termed the \textit{coarse propagator}, and $\mathcal{F}(t_{n},\mathbf{U}_{n})$, termed the \textit{fine propagator}.
For the first time step, the coarse propagator provides a coarse approximation to the solution with the initial condition $\mathbf{U}(t_{0}) = \mathbf{U}_{0}$, while the fine operator provides $\mathbf{U}(t_{1})$ with the desired accuracy.

We begin with some initial approximation, $\mathbf{U}_{n}^{0}$, for $n = \{0,1,2,\ldots,N\}$ corresponding to times $t_{n}$. This approximation is found by the application, in series, of the coarse propagator, i.e.:
\begin{equation}
	\mathbf{U}_{n+1}^{0} = \mathcal{G}(t_{n},\mathbf{U}_{n}^{0})\text{,   }\mathbf{U}_{0}^{0} = \mathbf{U}_{0}.
\label{eq:pint_setup}
\end{equation}
We then apply the \textit{correction iteration} for $k=0,1,2,\ldots$:
\begin{equation}
	\mathbf{U}_{n+1}^{k+1} = \mathcal{G}(t_{n},\mathbf{U}_{n}^{k+1}) + \mathcal{F}(t_{n},\mathbf{U}_{n}^{k}) - \mathcal{G}(t_{n},\mathbf{U}_{n}^{k}).
	\label{eq:pint_algorithm}
\end{equation}
We shall term equation~\eqref{eq:pint_algorithm} the \textit{Parareal} algorithm.
We note that as $k \to \infty$, it converges to:
\begin{equation}
\mathbf{U}_{N} = \hat{\mathcal{F}}(t_{N},t_{0},\mathbf{U}_{0})
\end{equation}
with $\hat{\mathcal{F}}$ computing the fine time steps from $t_0$ to $t_{N}$.
That is to say, the Parareal algorithm converges to the accuracy of the fine propagator. It has been proposed that the use of the Parareal algorithm permits this level of accuracy to be achieved more quickly in terms of wallclock time. Only the first step (equation \eqref{eq:pint_setup}) must be performed sequentially in time. For equation \eqref{eq:pint_algorithm}, there is no requirement of serial time and so processors which would otherwise be unused may now be used to refine the approximation.

It is worth noting that the fine propagator must solve the governing equation fully and to the desired accuracy.
Several different approaches have been taken to the coarse propagator, however.
$\mathcal{G}$ may in practice derive from a coarser timestep (e.g.\,\cite{lions2001resolution}), a coarser space discretisation (e.g.\,\cite{Fischer_etal_03}), a simpler physical model (e.g.\,\cite{Maday_Turinici_03}) and an exponential integrator (e.g.\,\cite{gander2013paraexp}) on which we further focus on.


\section{Decentralized Parareal}
\label{sec:decentralized_pint}

We introduce a software approach to allow a reutilization of the software for implementations of different kind of solvers and we present the required interfaces in Section \ref{sec:simulation_layer}.
The Parareal controller implements the logic behind the decentralized Parareal implementation and is presented in Section \ref{sec:Parareal_controller}.

%
%
%
%
%

\subsection{Simulation layer}
\label{sec:simulation_layer}

Each rank executes one instance of the simulation for a given coarse time interval.
With the Parareal algorithm given in its generic form (see \cite{lions2001resolution} and Section \ref{sec:parareal_algorithm}), the MPI parallelization can be hidden from the simulation developer.
In this Section, we describe the required interfaces with a focus on making the parallelization-in-time via MPI transparent to the simulation developer.
Please note that for sake of clarity, we skip the description of debugging and plotting features.
We group the interfaces in three different types: Setup, time stepping and Parareal difference/correction.
Several buffers are used and are denoted by $u^{\{S,F,C,D,O\}}$ (Start, Fine, Coarse, Difference, Output).


\subsubsection{Setup}
The \emph{setup} routines either depend on the initial conditions at $t:=0$ or 
simulation data forwarded by a previous coarse time frame.
\begin{itemize}
	\item \textit{constructor()}:\newline
		Constructor method for one-time-only initialization of the simulation instance.
	\item \textit{setSimulationTimeframe($t_{start}$, $t_{end}$)}:\newline
		Set the time frame for the coarse time interval.
	\item \textit{setupInitialValues()}:\newline
		Setup the initial values at $t=0$
	\item \textit{setSimulationData(data)}:\newline
		Set simulation data $u_S := data$.
\end{itemize}

The constructor initializes the simulation only once for each rank.
This allows an efficient sliding window by only requiring to set the new simulation time frame via \textit{setSimulationTimeframe} and by the new initial values via \textit{setSimulationData} without requiring reinitializing e.g.\,FFT computations.

\subsubsection{Timestepping}
The timestepping interfaces are required to execute the fine and coarse timesteps.
The results of these time stepping methods are then made available via \textit{get}ters.

\begin{itemize}
	\item \textit{runTimestepFine()}\newline
		Compute the solution at $t_{end}$ with the fine timestepping method:
		$u^F := \mathcal{F}(t_{end}, t_{start}, u^S)$
	\item \textit{runTimestepCoarse()}\newline
		Compute the solution at $t_{end}$ with the coarse timestepping method:
		$u^C := \mathcal{G}(t_{end}, t_{start}, u^S)$
	\item \textit{getDataTimestepFine()}\newline
		Return the solution $u^F$
	\item \textit{getDataTimestepCoarse()}\newline
		Return the solution $u^C$
\end{itemize}

\subsubsection{Parareal difference/correction}
Finally, the solutions of the different timestep methods have to be merged together (see Eq.\,\eqref{eq:pint_algorithm}) in a certain way without race conditions which can be accomplished by the following interfaces:
\begin{itemize}
	\item \textit{computeDifference()}\newline
		Compute the difference between the fine and coarse time stepping $u^D := u^F - u^C$.

	\item \textit{computeOutputData()}\newline
		Compute the data to be forwarded to the next timestep by applying a correction to the solution from the coarse timestep: $u^O := u^C + u^D$.
		Note, that $u^C$ is based on the coarse timestep executed \emph{after} calling \textit{computeDifference}.

	\item \textit{getOutputData()}\newline
		Return the reference to the data $u^O$ to be forwarded to the next coarse timestep interval.

	\item \textit{getErrorEstimation()}\newline
		Return a scalar value as an error estimation.
		This is typically based on a norm of the computed solution and is required for the convergence test.
\end{itemize}

These interfaces contain no information on the adjacent MPI ranks and hide the connectivity from the simulation developer.
Next, we discuss the logic which triggers the execution of these interfaces and which orchestrates several coarse timestep intervals.

\subsection{Parareal controller}
\label{sec:Parareal_controller}

\begin{figure}
	\center
	\includegraphics[width=0.5\textwidth]{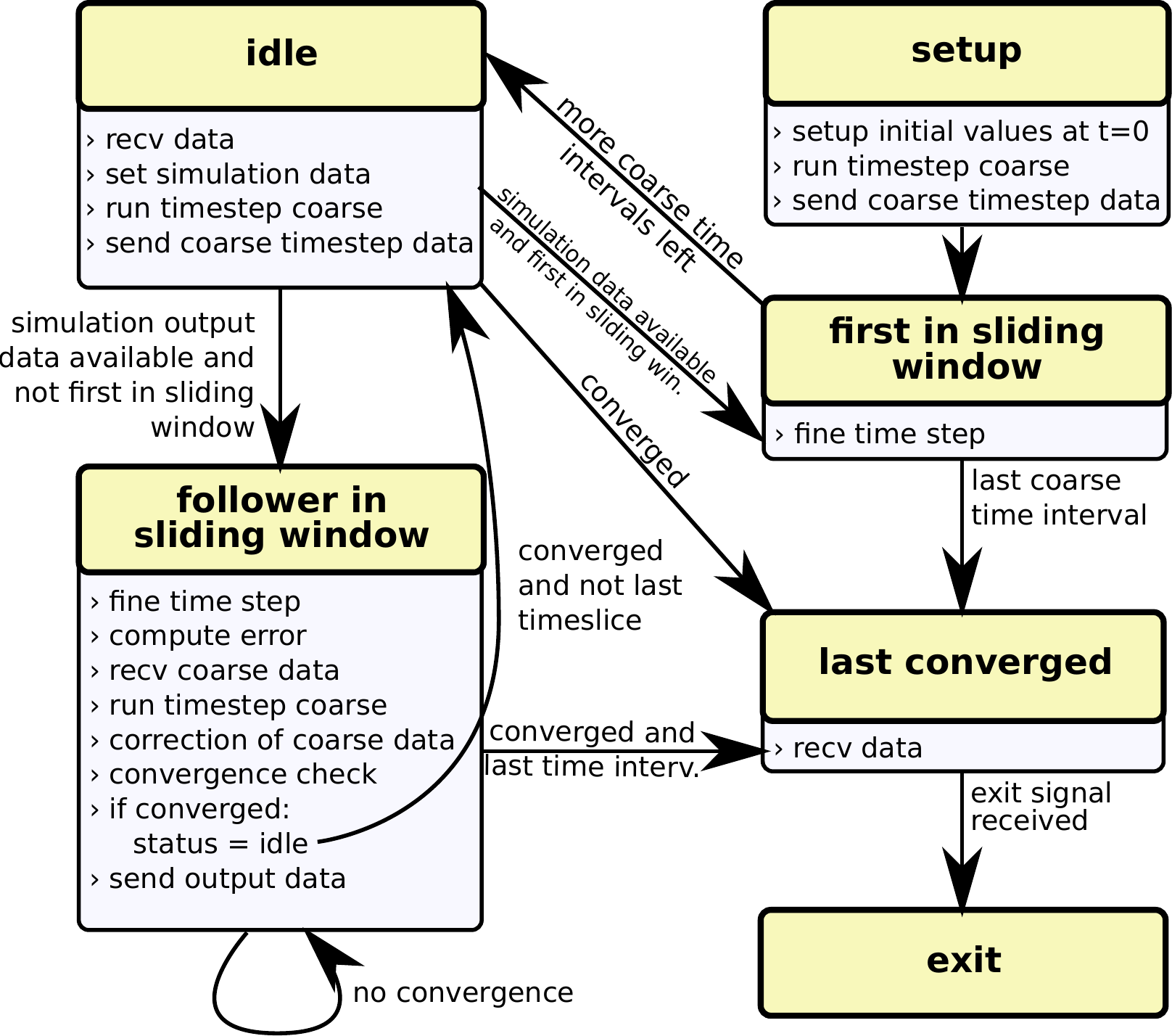}
	\caption{
		Overview of the different states of the Parareal controller simulation instances.
		Each box represents one of the six states.
		A short description of the most important operations executed for each state is given in below each state box.
		The transitions depend on the convergence or state behaviour.
		The receive/send operations are done from/to the previous/next ranks only.
		\label{fig:time_slice_automaton_states}
	}
\end{figure}

The Parareal controller implements the entire logic behind our decentralized Parareal approach.
It is mainly based on a state machine with the transitions depending on (a) the current state, (b) the state information forwarded by the previous rank and (c) the convergence test.
After initialization, the first rank is set to the \textit{[setup]} state and all other ranks to the \textit{[idle]} state.
All possible states are discussed in more detail in the following list and an overview is given in Fig.\,\ref{fig:time_slice_automaton_states}.

\begin{itemize}
	\item \textbf{[setup]}\newline
		The first rank $i = 0$ is set to the [setup] state.
		This triggers the setup of the initial values at t=0.
		Then, a coarse timestep is computed and the data forwarded to rank $1$.
		After this setup, the state changes to [first in sliding window].

	\item \textbf{[first in sliding window]}\newline
		A fine timestep is executed.
		Since this is the first coarse time interval in the sliding window, further Parareal iterations would not yield an improvement in the solution.
		Therefore, the solution of the just computed fine timestep is forwarded to the next rank and the state is set to idle.

	\item \textbf{[follower in sliding window]}\newline
		A follower in the sliding window first executes the \emph{fine timesteps} (runTimestepFine).
		Then, the \emph{difference} between the solution of the coarse and fine timesteps are computed (computeDifference).
		This is followed by waiting for new simulation data at $t_{start}$ from the previous rank which is then
		used for executing a coarse timestep (runTimestepCoarse).
		The solution of the coarse timestep is then corrected by the previously computed difference
		and the result forwarded to the next rank (computeOutputData).

		The state change depends on the previous coarse time interval:
		In case of \emph{no convergence of the previous coarse time interval}, the \textit{state is unchanged}.
		With a \emph{convergence in the previous coarse time interval and the current one}, the state is changed to \textit{[idle]}.
		\emph{Otherwise}, the state of the coarse time interval becomes the \textit{[first in the sliding window]}.

	\item \textbf{[idle]}\newline
		An idling state checks for messages from previous ranks.
		Due to our asynchronous and decentralized approach, it is possible that more than one simulation data states are already enqueued in the receive buffer.
		Therefore, we probe for such additional messages and in the case that new simulation data is already available, we drop the previous one and read this next message.
		Depending on the state of the previous rank, the state is changed to \textit{[follower in sliding window]} or \textit{[first in sliding window]}.
		In case of receiving a converged state from the previous rank, the state is changed to \textit{[last converged]}.

	\item \textbf{[last converged]}\newline
		This state can be only reached if the last coarse time interval in the entire simulation time frame was reached.
		A transition to this state is either triggered via the first/follower in case of a convergence of the last coarse time interval or by receiving this state by the previous coarse time interval (see transition from \textit{[idle]}).
		During this state, messages from previous ranks are still received to assure that no network congestion occurs.
		After transition to this state, the \textit{[exit]} state is send to the next rank who can receive this message only, if all other simulation data messages were read.


	\item \textbf{[exit]}\newline
		With the algorithm presented in \textit{[last converged]}, a transition to \textit{[exit]} is done if receiving the \textit{[exit]} signal.
		After assuring that all messages were send in the sending queue, this instance of the coarse time interval of the Parareal simulation exits.
\end{itemize}

%



\section{Results}
\label{sec:results}

We conducted several studies based on an experimental implementation of the rotational shallow water equations in combination with our decentralized parareal parallelisation (Sec.\,\ref{sec:decentralized_pint}).
These studies were focused on a particular set of parameters which are set as follows:
The benchmarks were conducted with different resolutions $r \in \{8^2, 16^2, 32^2, 64^2, 128^2\}$ for the simulation domain.
We use a fine time step size of $0.001$ which is sufficiently small for stability reasons for all values of $r$ regarding the CFL condition and by using an exponential integrator for the linear part.
For the Parareal method, we use a coarse time step size of $0.1$, hence we execute 100 fine time step sizes within a single coarse time step.
We use a Gaussian distribution $\frac{1}{2} \exp{-5 ((x-\pi)^2 + (y-\pi)^2)}$ for the initial values.
The simulation is executed over 40 seconds of wave propagations.
The convergence test is based on the data which is forwarded to the next coarse time interval.
Here, we use the minimum of the $L_2$ and $L_{max}$ norm and set the threshold for the convergence test to $10^{-5}$.
For sake of reproducibility, the source code for the Parareal framework is available for download \cite{pypint_release}.
The blackbox RSWE solver can be requested from the 2nd author.

\begin{figure*}
	\center
	\includegraphics[width=0.75\textwidth]{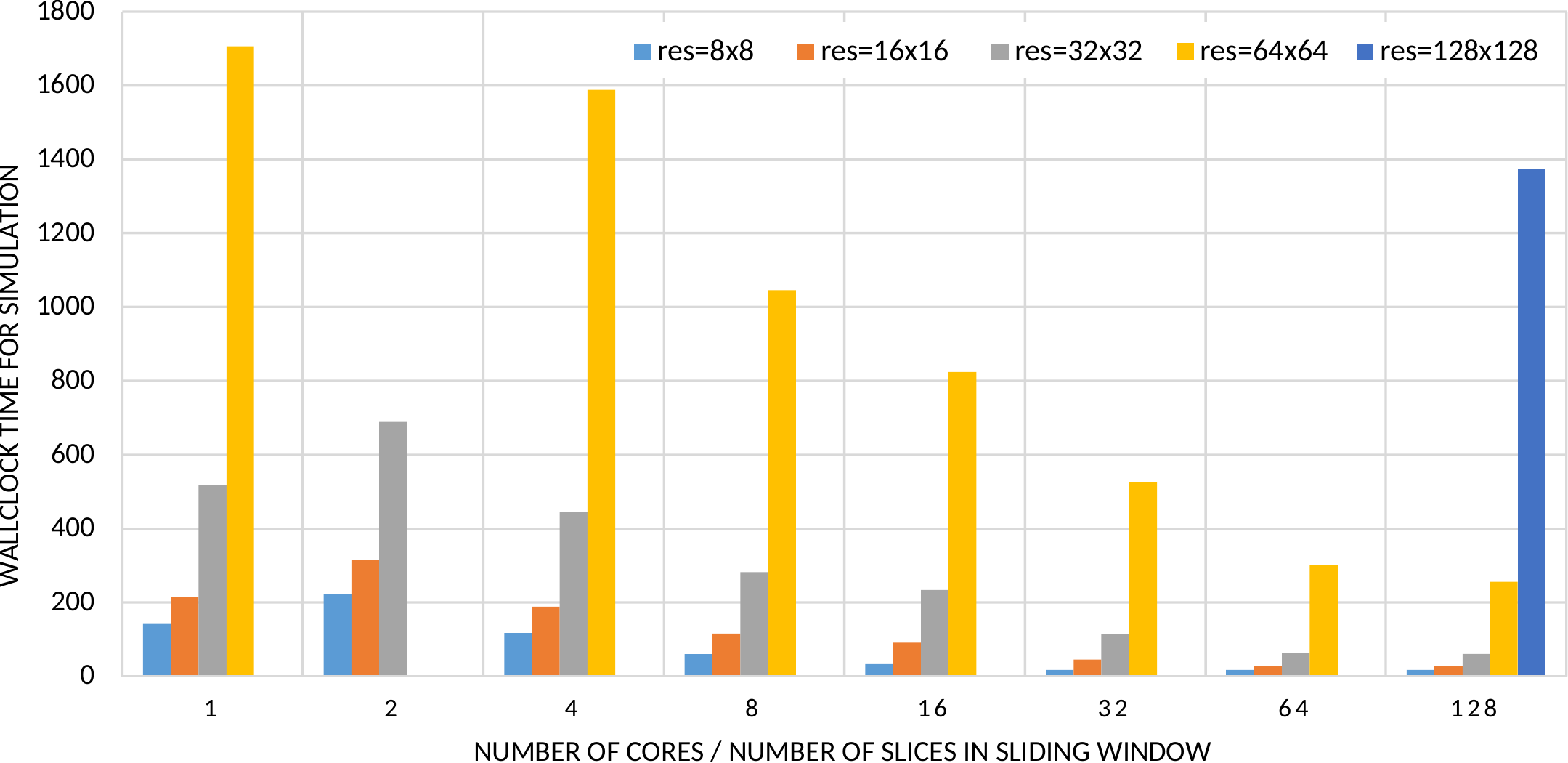}
	\caption{
		Wallclock time for solving the RSWE with different resolutions.
		Since we focus on problems with a limitation on the time-to-solution, we plotted the results with the wallclock time in the y-axis for a better comparison.
		The wallclock time for a single core was computed with the fine timestepping only and the overall runtime was restricted to 30 minutes.
		For resolution $64^2$, we see an increase of runtime with two cores.
		This is due to additional computational costs of the coarse time step and communication overheads.
		For a higher number of cores, there is a robust decrease of computation time with a speedup of $6.7$ for using 128 cores.
		A simulation with the resolution of $128^2$ (blue right-most bar) only gets feasible by using 128 cores with the time restriction of 30 minutes.
		\label{fig:runs_threaded_ssend_eager_100ftspcts}
	}
\end{figure*}

We conducted scalability benchmarks for up to 128 cores with the results given in Fig.\,\ref{fig:runs_threaded_ssend_eager_100ftspcts}.
Regarding a parallelization-in-space, we expect that there will be no scalability possible across several MPI compute nodes for the considered problem sizes.
Indeed, a scalability even on shared-memory many-core systems which would not suffer of MPI communication overheads is hardly feasible due to a \emph{two-dimensional} problem of about $64 \times 64$.
Here, the non-parallelizable parts in the parallel execution (threading overheads, cache-synchronisation, NUMA effects, bandwidth limitation, etc.) would dominate and significantly restrict the scalability already on such shared-memory systems.
Therefore, we focus in the following on a parallelization-in-time only.

The runtime was restricted to 30 minutes to account for real-time requirements which was the original motivation of the Parareal approach.
For the single-core performance, we only used the fine time stepping method without any communication and Parareal overheads.
This performance is used as the baseline in the following.
We can see an increase of wallclock time for executing the simulation on two cores.
We account for that by (a) the additional time required for the coarse time stepping, (b) the communication overheads of sending the simulation data to the next MPI rank and (c) the convergence test which requires at least two iterations.
In particular because of issue (c), there cannot be any performance increase of the Parareal method with only two coarse time steps and by using a convergence test.
By using four coarse time intervals in the sliding window, we already gain a robust speedup for all considered resolutions.
Utilizing $128$ cores for the parallelization-in-time, we get speedups of $(8.64, 7.74, 8.54, 6.66)$ for the resolutions $(8^2, 16^2, 32^2, 64^2)$, respectively.


\bibliography{paper_ver_2016_02}

\begin{thebibliography}{10}
\providecommand{\url}[1]{#1}
\csname url@samestyle\endcsname
\providecommand{\newblock}{\relax}
\providecommand{\bibinfo}[2]{#2}
\providecommand{\BIBentrySTDinterwordspacing}{\spaceskip=0pt\relax}
\providecommand{\BIBentryALTinterwordstretchfactor}{4}
\providecommand{\BIBentryALTinterwordspacing}{\spaceskip=\fontdimen2\font plus
\BIBentryALTinterwordstretchfactor\fontdimen3\font minus
  \fontdimen4\font\relax}
\providecommand{\BIBforeignlanguage}[2]{{%
\expandafter\ifx\csname l@#1\endcsname\relax
\typeout{** WARNING: IEEEtran.bst: No hyphenation pattern has been}%
\typeout{** loaded for the language `#1'. Using the pattern for}%
\typeout{** the default language instead.}%
\else
\language=\csname l@#1\endcsname
\fi
#2}}
\providecommand{\BIBdecl}{\relax}
\BIBdecl

\bibitem{kogge2011using}
P.~M. Kogge and T.~J. Dysart, ``Using the top500 to trace and project
  technology and architecture trends,'' in \emph{Proceedings of 2011
  International Conference for High Performance Computing, Networking, Storage
  and Analysis}.\hskip 1em plus 0.5em minus 0.4em\relax ACM, 2011, p.~28.

\bibitem{lions2001resolution}
J.-L. Lions, Y.~Maday, and G.~Turinici, ``R{\'e}solution d'edp par un
  sch{\'e}ma en temps parar{\'e}el,'' \emph{Comptes Rendus de l'Acad{\'e}mie
  des Sciences-Series I-Mathematics}, vol. 332, no.~7, pp. 661--668, 2001.

\bibitem{elwasif2011dependency}
W.~R. Elwasif, S.~S. Foley, D.~E. Bernholdt, L.~A. Berry, D.~Samaddar, D.~E.
  Newman, and R.~Sanchez, ``A dependency-driven formulation of parareal:
  parallel-in-time solution of pdes as a many-task application,'' in
  \emph{Proceedings of the 2011 ACM international workshop on Many task
  computing on grids and supercomputers}.\hskip 1em plus 0.5em minus
  0.4em\relax ACM, 2011, pp. 15--24.

\bibitem{Aubanel2011172}
\BIBentryALTinterwordspacing
E.~Aubanel, ``Scheduling of tasks in the parareal algorithm,'' \emph{Parallel
  Computing}, vol.~37, no.~3, pp. 172 -- 182, 2011. [Online]. Available:
  \url{http://www.sciencedirect.com/science/article/pii/S0167819110001419}
\BIBentrySTDinterwordspacing

\bibitem{Nievergelt_64}
J.~Nievergelt, ``Parallel methods for integrating ordinary differential
  equations,'' \emph{Commun. ACM}, vol.~7, no.~12, pp. 731--733, Dec. 1964.

\bibitem{Maday_Turinici_05}
Y.~Maday and G.~Turinici, ``The parareal in time iterative solver: a further
  direction to parallel implementation,'' in \emph{Domain decomposition methods
  in science and engineering}.\hskip 1em plus 0.5em minus 0.4em\relax Springer
  Berlin Heidelberg, 2005, pp. 441--448.

\bibitem{Gander_15}
M.~J. Gander, ``50 years of time parallel time integration,'' in \emph{Multiple
  Shooting and Time Domain Decomposition}, T.~Carraro, M.~Geiger, S.~Korkel,
  and R.~Rannacher, Eds.\hskip 1em plus 0.5em minus 0.4em\relax
  Springer-Verlag, 2015.

\bibitem{Fischer_etal_03}
P.~F. Fischer, F.~Hecht, and Y.~Maday, ``\BIBforeignlanguage{English}{A
  parareal in time semi-implicit approximation of the navier-stokes
  equations},'' in \emph{\BIBforeignlanguage{English}{Domain Decomposition
  Methods in Science and Engineering}}, ser. Lecture Notes in Computational
  Science and Engineering, T.~J. Barth, M.~Griebel, D.~E. Keyes, R.~M.
  Nieminen, D.~Roose, T.~Schlick, R.~Kornhuber, R.~Hoppe, J.~Périaux,
  O.~Pironneau, O.~Widlund, and J.~Xu, Eds.\hskip 1em plus 0.5em minus
  0.4em\relax Springer Berlin Heidelberg, 2005, vol.~40, pp. 433--440.

\bibitem{Maday_Turinici_03}
Y.~Maday and G.~Turinici, ``Parallel in time algorithms for quantum control:
  Parareal time discretization scheme,'' \emph{International journal of quantum
  chemistry}, vol.~93, no.~3, pp. 223--228, 2003.

\bibitem{gander2013paraexp}
M.~J. Gander and S.~Güttel, ``Paraexp: A parallel integrator for linear
  initial-value problems,'' \emph{SIAM Journal on Scientific Computing},
  vol.~35, no.~2, pp. C123--C142, 2013.

\bibitem{pypint_release}
\BIBentryALTinterwordspacing
M.~Schreiber \emph{et~al.} [Online]. Available:
  \url{http://www.martin-schreiber.info/pub/exeter/pypint/release_pypint_2015_11_14.tar.bz2}
\BIBentrySTDinterwordspacing

\end{thebibliography}

\end{document}